\documentstyle[prl,aps,epsfig]{revtex}

\input{psfig}

\begin{document}

\title{Sound
attenuation on the Bose-Einstein condensation of magnons in $TlCuCl_3$}

\author{E. Ya. Sherman$^1$, P. Lemmens$^{2,3}$,
B. Busse$^3$, A. Oosawa$^4$, H. Tanaka$^4$}

\address{$^1$Institute for Theoretical Physics,
       Karl-Franzens-University of Graz, A-8010, Graz, Austria}
\address{$^2$ Max-Planck-Institute for Solid State Research,
   D-70569 Stuttgart, Germany}
\address{$^3$ IMNF and High-Field Laboratory, TU Braunschweig, D-38106 Braunschweig,
Germany}
\address{$^4$ Department of Physics, Tokyo
Institute of Technology, Oh-okayama, Meguro-ku, Tokyo 152-8551, Japan}

\date{June 24th, 2003}

\maketitle
\begin{abstract}

We investigated experimentally and theoretically sound attenuation in the
quantum spin system $TlCuCl_{3}$ in magnetic fields at low temperatures.
Near the point of Bose-Einstein condensation (BEC) of magnons a sharp peak
in the sound attenuation is observed. The peak demonstrates a hysteresis as
function of the magnetic field pointing to a first-order contribution to the
transition. The sound damping has a Drude-like form arising as a result of
hardcore magnon-magnon collisions. The strength of the coupling between
lattice and magnons is estimated from the experimental data. The puzzling
relationship between the transition temperature and the concentration of
magnons is explained by their "relativistic" dispersion.

\end{abstract}
% indicates a first-order contribution to the character of the transition
% change abstract


Due to the richness and universality of phenomena that may be observed
\cite{Auerbach} quantum spin systems in magnetic fields have recently gained
enormous attention. Similarities of quantum spin systems \cite{Lemmens03}
and quantum fields allow to test experimentally and develop predictions of
quantum field theories. The observation of plateaus in the high-field
magnetization of one-dimensional quantum spin systems that follow the
Oshikawa-Yamanaka-Affleck condition \cite{oshikawa97} is an example of this
mutual interest. The effect of a magnetic field on a gapped spin system is
especially interesting for a small gap, which can be reduced or even closed
by experimentally available static fields. On the other hand sound
attenuation experiments were very useful in the investigation of phase
transitions with complex order parameters, e.g. superconductivity in the
heavy fermion $\rm UPt_3$ \cite{ultrasound} or the movement of flux lines in
high-$T_c$ superconductors \cite{pankert90}.

The $\mathrm{XCuCl_{3}}$ family of compounds, with X= Tl, K and
$\mathrm{NH_{4}}$, realize especially interesting systems \cite{Lemmens03}.
Magnetization plateaus are observed in $\mathrm{NH_{4}CuCl_{3}}$, while
$\mathrm{TlCuCl_{3}}$ and $\mathrm{KCuCl_{3}}$ show field-induced critical
phenomena. The underlying quantum magnetism is based on  $S=1/2$ spins of
$\mathrm{Cu^{2+}}$ ions arranged as planar dimers of
$\mathrm{Cu_{2}Cl_{6}}$. These dimers form infinite double chains parallel
to the crystallographic $a$-axis of the monoclinic structure (space group
$\mathrm{P2_{1}/c}$ with four formula units per unit cell) \cite
{willett63,tanaka96,takatsu97}. The ground state of $\mathrm{TlCuCl_{3}}$ is
a singlet with a spin gap $\mathrm{\Delta \approx 7~K}$ as a weakly
anisotropic antiferromagnetic intra-dimer interaction of $J\sim $60~K
exists. The magnon dispersion of this compound was investigated
theoretically in Ref. \cite{Matsumoto02}. The three-dimensional (3D)
character of the system leads to a relatively small gap since in this case
it is more difficult to bind the magnons. In a magnetic field $H$ the gap
for $S_{z}=-1$ excitations is reduced as $\Delta-g\mu_{\rm B}H$ where a $g-$
factor varies between 2.06 and 2.23 depending on the ${\bf H}$- direction.
Reaching a field of $H_{g}=\Delta /g\mu\mathrm{_{B}}$, the gap is closed.

A possible mechanism of magnetic-field induced spin transitions introduced
by Giamarchi and Tsvelik \cite{Giamarchi99} is the Bose condensation of the
soft mode.  For the explanation of field-induced magnetic ordering in
$\mathrm{TlCuCl_{3}}$ the Bose condensation of diluted magnons with
$S_{z}=-1$ was proposed \cite{nikuni00}. A Hartree-Fock description of such
a transition \cite{nikuni00} leads to an exponent $\phi $=1.5 relating
concentration of magnons $n_{B}$ and the transition temperature $T_{B}$ as
$n_{B}\sim T_{B}^{\phi }$. However, a larger value of $\mathrm{\phi _{exp}}$
close to 2 has been determined by magnetization and specific heat
experiments \cite{nikuni00,oosawa99,oosawa01,tanaka01}.

At the same time, the Bose condensation should be observable in experiments
which are directly related to the distribution function of the magnons
$f_B(k)$, where $k$ is the magnon momentum. Here we present, to our
knowledge, the first results of ultrasonic attenuation $\alpha (H)$
measurements on $\mathrm{TlCuCl_{3}}$ in magnetic field and show that the
experimental data suggests a gradual increase of $f_B(k)$ in the region of
small $k$ with the increase of $H$ being consistent with the BEC of magnons
at a critical field $H_{c}.$ As we will see below, ultrasonic attenuation
experiments in this compound allow to trace the main features of the
distribution function, investigate the fluctuation region where the
attenuation anomalously increases, make a conclusion on the order of the
transition, and determine the mechanism and strength of magnon-lattice
coupling.

Ultrasonic attenuation has been measured in single crystals
of $\mathrm{TlCuCl_{3}}$ prepared by the Bridgman method \cite{oosawa99} using a
pulse-echo method with longitudinal sound waves at a frequency of 5 MHz. The
transducers were glued on freshly cleaved or on polished surfaces in the
crystallographic $(10\overline{2})$ plane of the crystals using a liquid
polymer (Thiokol, Dow Resin). The ${\bf H}$-field has been applied parallel
to the direction of sound propagation. The value of $\alpha (H)$ is
determined from the experimental data as $\alpha (H)=10\lg (I_{N}/I_{N+1}),$
where $I_{N}$ and $I_{N+1}$ are the intensities of $N$th and $N$th$+1$
pulses arriving the detector with the time interval $2t_{p},$ where $t_{p}$
is the pulse travelling time through the sample.

%%%%%%%%%%%%%%%%%%%%%%%%%%
 \begin{figure}
\centerline{\epsfig{figure=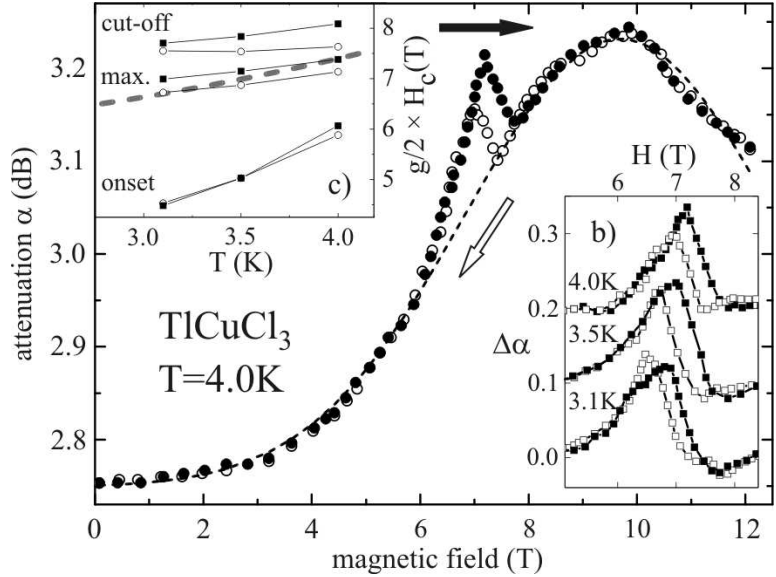,width=11cm}} \caption{Ultrasonic
attenuation $\alpha(H)$ with increasing (full circles) and decreasing
magnetic field (open circles) and a fitting of the broad maximum by a
Gaussian (dashed line). Inset (b): Transition regime after subtracting the
Gaussian and a constant. Inset (c): Phase diagram with ${gH_c/2}$ (thick
line) from thermodynamic experiments \protect\cite{nikuni00}. The $g$-factor
of $S_z=-1$ excitations $g=2.16$.} \label{fig1}
 \end{figure}
%%%%%%%%%%%%%%%%%%%%%%%%%%

The experimental results shown in Fig.~1 can be summarized as the
following observations:

1) a weak $H$-dependence up to some upturn at $H\approx 2.5$ T,

2) a sharp and asymmetric peak close to the critical field $H\mathrm{_{c}}$
determined by thermodynamic experiments \cite {nikuni00}. The sharp
peak associated with the fluctuation region of the transition shows a
pronounced hysteresis with increasing/decreasing field, which
indicates a first-order contribution to the transition, and

3) a decrease of $\alpha(H)$ for $H>9.7$ T.

If the comparably slow underlying increase is approximated by a Gaussian,
the sharp peak near $H_{c}$ can be separated from the background.
Inset (b) shows the resulting transition-induced attenuation at three
different temperatures. The peaks
have a gradual low-field onset of attenuation ${H_{c}^{%
\mathrm{ons}}}$, a sharp maximum ${H_{c}^{\mathrm{max}}}$ and a high-field
cut-off $H_{c}^{\mathrm{co}}$. The two critical fields
${H_{c}^{\mathrm{max}}}$ and $H_{c}^{\mathrm{co}}$ show a pronounced
hysteresis. The respective attenuation at the critical fields systematically
changes with $T$. For ${H_{c}^{\mathrm{ons}}}$, that marks
a crossover into a regime with strong fluctuations, the hysteresis is
negligible. In a recent NMR investigation a hysteresis close to the
phase boundary have also been observed and attributed to spin-phonon
coupling \cite{takigawa01}. Inset (c) shows the resulting phase diagram
compared to magnetization measurements \cite{nikuni00}\ (thick dashed line).

The energy of a propagating sound pulse decays with time as: $w(t)\sim \exp
[-(\Gamma +\gamma _{m}(H))t]$, and, therefore $\alpha (H)=20\lg e\cdot
\left[ \Gamma +\gamma _{m}(H)\right] t_{p}.$ Here $\gamma _{m}(H)$ is the
contribution of magnons coupled to the lattice, while an unknown background
$\Gamma $ arises due to all other effects like scattering by impurities,
interfaces, etc. In contrast to $\Gamma $, the $H$-dependence of the damping
$\gamma _{m}(H)-\gamma _{m}(0)$ which corresponds to the field-induced
changes in $f_B(k)$  and sound attenuation mechanism can be determined
accurately from Fig.~1. By comparison with theoretical calculations also
$\gamma _{m}(0)$ will be obtained from the experiment.

Cottam \cite{Cottam74} investigated spin-phonon coupling in an
antiferromagnet in the case of a low density of magnons where magnon-magnon
interactions can be neglected. Below we investigated another case where the
damping is related to collisions between magnons. The model for sound
attenuation is the following: In the field of the longitudinal sound wave
with displacements of ions ${\mathbf u}({\mathbf r})$, the energy of the
magnon changes due to magnon-lattice coupling as $\delta \varepsilon
({\mathbf k})=C_{k}\mathrm{\nabla }{\mathbf u}({\mathbf r})$, where $C_{k}$
is the deformation potential.  The spatial dependence of the displacement in
a plane wave with wavevector $q$ and frequency $\omega$:  ${\mathbf
u}({\mathbf r})={\mathbf u}_{0}\exp [i({\mathbf q}{\mathbf r}-\omega t)]$
leads to an effective time-dependent force ${\mathbf F}=-\nabla \delta
\varepsilon({\mathbf k}) $ with $F=C_{k}u_{0}q^{2}$ driving changes in the
velocity of magnons. Since one cannot create two $S_{z}=-1$ states on a
Cu-Cu bond, the magnons collide like hard-core bosons. Inelastic collisions
between the magnons driven by the displacements of the lattice in the wave
lead to a dissipation of the pulse energy. The dissipation rate per one
magnon is equal to the average of the product ${\mathbf F}{\mathbf v}$ per
period, where ${\mathbf v}$ is the magnon velocity. The sound dissipation
rate in this case is the dissipation rate of an external harmonic field in a
medium, generally described by a Drude-like mechanism of the form:
\begin{equation}
\hspace{-2cm}\gamma _{m}(H)= \frac{Au_{0}^{2}q^{4}}{(2\pi)^{6}} \frac{\left(
n\sigma \right) ^{-1}}{1+\omega^{2}\langle\tau\rangle^{2}} \int
{d^{3}k}f_{B}(k)\frac{C_{k}^{2}}{m_{k}} \int{d^{3}k^{\prime }}
f_{B}(k^{\prime }) \frac{1}{nv_{{\mathbf k},{\mathbf k}^{\prime}}},
\end{equation}
where $m_{k}$ is the magnon effective mass and $v_{{\mathbf k},{\mathbf k}^{\prime }}$ is the
relative velocity of the particles with momenta ${\mathbf k}$ and ${\mathbf
k^{\prime }}$ that determines their collision rate. The constant $A$ is of
the order of one and depends on details of magnon-magnon collisions, which
are beyond our consideration. The mean time between the
collisions $\langle\tau\rangle\sim\langle\ell v^{-1}\rangle$ with the magnon
free path $\ell $ being proportional to $1/n\sigma $, where $\sigma$ is the
magnon-magnon scattering cross-section, and $n$ is the magnon concentration,
is, therefore sensitive to the distribution function. The mean value
$\langle v^{-1}\rangle $ and the damping increase simultaneously. Since
$\omega \langle \tau\rangle $ is very small $\left( \sim 0.01\right)$ at the
experimental $\omega \sim 3\cdot 10^{7}$ s$^{-1}$, it will be
neglected in the denominator of Eq.(1) for the rest of the discussion.

To include the specific spin excitation spectrum, we start
with the ''relativistic'' form of the magnon dispersion $\varepsilon
(k)=\sqrt{\Delta ^{2}+J^{2}k^{2}/4}$ at small $k$ which is important if
$T_{B}$ and $\Delta $ are of the same order of magnitude as in
$\mathrm{TlCuCl_{3}}$. From here on the lattice constant $l\equiv 1$ if
otherwise is not stated. Since thermal $k\ll 1$ at $T_{B}\ll J$  this form
of energy is sufficient for an understanding of the Bose condensation in the
experimental temperature range. The velocity of a magnon and its inverse
mass are (we put $\hbar\equiv 1$):
\begin{equation}
v(k)=\frac{d\varepsilon (k)}{dk}=\frac{J^{2}}{4\varepsilon (k)}k,\qquad
\frac{1}{m_{k}}=\frac{J^{2}}{4}\left[ \frac{1}{\varepsilon
(k)}-\frac{J^{2}}{%
4}\frac{k^{2}}{\varepsilon ^{3}(k)}\right] ,
\end{equation}
respectively. The mass at small $k$, $m={4\Delta }/{ J^{2}}$, determines
$T_B$  at low concentration of magnons where $T_{B}\ll \Delta$. In the
mean-field approximation which will be used below, the distribution
function is $f_B(\varepsilon )=1/[\exp((\overline{\varepsilon }-\mu
_{\mathrm{eff}})/T)-1]$, where $\overline{\varepsilon}(k)=\varepsilon
(k)-\Delta$, and $\mu_{\mathrm{eff}}=g\mu_{B}(H-H_{g})-2V_{\rm
mm}\overline{n}$. $\overline{n}$ is the dimensionless concentration of the
magnons determined by the
magnetization $M$ as $\overline{n}%
=M/H$, and $V_{\rm mm}$ is the magnon-magnon interaction parameter. The transition temperature
$T_{B}$ obtained with the
condition $\mu _{\mathrm{eff}}=0$, that is
\begin{equation}
4\pi \int \frac{k^{2}}{e^{\overline{\varepsilon}(k)/T_{B}}-1}\frac{dk}{\left( 2\pi
\right)^{3}}=n_{B},
\end{equation}
is, therefore, $V_{\rm mm}$-independent. Fig.2a presents dimensionless
$\overline{n}_{B}(T_{B})$, and Fig.2b shows the ratio
$\left(\overline{n}_{B}(T_{B})/\overline{n}_{B}(1\mathrm{ K})\right)/T_{B}^{2}$ obtained
from Eq.(3) for different $\Delta$ and $J=7\Delta$. As one can see, at the
chosen $\Delta=6.5$ K, the ratio is nearly a
constant in the experimental range of temperatures, giving an explanation
for the observation of $\mathrm{\phi _{exp}}\approx 2$
\cite{nikuni00,oosawa99,oosawa01,tanaka01}. Since near the condensation
point the thermal wavevector of the magnon $%
k^{2}\sim mT_{B}$, leading to a temperature dependence of the mean value
$\langle \varepsilon (k)-\Delta \rangle \sim T$, the ``exponent'' depends on
$T_{B}/\Delta$ only. We note that $\phi_{\rm exp}$ is indeed a fitting
parameter, which can depend on the measured quantity. We also mention that
the parameter which determines the applicability of the Drude approach
$k\ell$ and can be estimated as $\sim\overline{n}^{-2/3}$, where $k$ is the
characteristic magnon momentum, is always larger than ten thus justifying
the validity of Eq.(1) except the fluctuation region.

 %%%%%%%%%%%%%%%%%%%
 \begin{figure}
 \centerline{\epsfig{figure=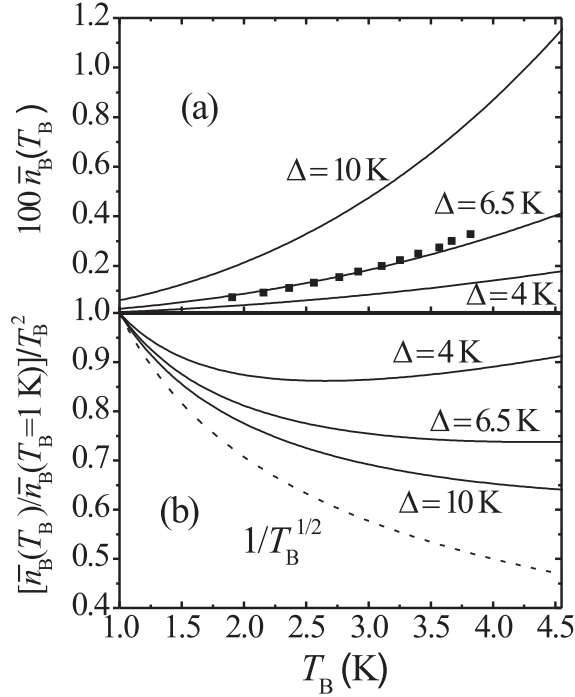,width=10cm}}
\caption{(a) Dimensionless concentration of magnons
$\overline{n}_{B}(T_{B})$ corresponding to the BEC temperature $T_{B}$.
Squares present experimental data \protect\cite{nikuni00}. (b)
$\left[\overline{n}_{B}(T_{B})/\overline{n}_{B}(1\mathrm{K})\right]T_{B}^{-2}$.
With increasing $\Delta$ the curves are closer to $T_{B}^{-1/2}$ (the dashed
line), corresponding to $\phi=3/2$.}
 \label{fig2}
 \end{figure}
 %%%%%%%%%%%%%%%%%%%%%%%%%%

% %%%%%%%%%%%%%%%%%%%
% \begin{figure}
% \centerline{\epsfig{figure=figure2.eps,width=10cm}}
%\caption{(a) Dimensionless concentration of magnons
%$\overline{n}_{B}(T_{B})$ corresponding to the BEC temperature $T_{B}$.
%Squares present experimental data \cite{nikuni00}. (b)
%$\left[\overline{n}_{B}(T_{B})/\overline{n}_{B}(1\mathrm{K})\right]T_{B}^{-2}$.
%With increasing $\Delta$ the curves are closer to $T_{B}^{-1/2}$ (the dashed
%line), corresponding to $\phi=3/2$.}
% \label{fig2}
% \end{figure}
% %%%%%%%%%%%%%%%%%%%%%%%%%%

Since near the transition the Bose function behaves as
\begin{equation}
f_{B}\left( \varepsilon \right) =\frac{T_B}{\overline{\varepsilon} +|\mu
_{\mathrm{eff}}|},
\end{equation}
due to the singularity at $\mu _{\mathrm{eff}}\rightarrow 0$, the average $%
\langle v^{-1}\rangle $ diverges as $\ln (T_{B}/|\mu _{\mathrm{eff}}|)$
leading, as it will be shown below, to an increase in the damping rate.

The sound-induced change in the magnon energy consists of two terms,
$\delta\varepsilon ({\mathbf k})=\delta \varepsilon_{\Delta }({\mathbf k})+\delta\varepsilon _{J}({\mathbf k})$
arising due to a modulation in the
gap $\widetilde{\Delta }=\Delta+C_{\Delta}{\mathrm \nabla}{\mathbf u}$ and
the exchange $\widetilde{J}=J+C_{J}\mathrm{\nabla}{\mathbf u}$,
respectively.
The resulting deformation potential is:
\begin{equation}
C_{k}=\frac{\Delta }{\varepsilon (k)}C_{\Delta}+\frac{J}{\varepsilon (k)}\frac{k^{2}}{4}C_{J}.
\end{equation}
Since the $J$-originated term in $C_{k}$ vanishes at small $k$, the main
contribution to the increase of damping near the transition
comes from $C_{\Delta }$. The calculated behavior of the
damping at $T$=3.5 K for $H<H_{c}$
is shown in Fig. 3 for two $V_{\rm mm}$ and two models of
phonon-magnon coupling. As a result of the increase of
$f_{B}(k)$ at small $k$ near the transition point, the damping due to the
$C_{\Delta}$-term increases and shows a $\gamma _{m}(H)$ behavior which is
in full agreement with the experimental data in Fig. 1. At the same time,
due to a vanishing coupling for $k\rightarrow 0$ in the second
contribution in Eq.(5), the $C_{J}-$dependent term leads to a slight
decrease in the damping, in contrast to Fig.1. Moreover, near the
transition point, where $k^{2}\sim T_{B}\Delta /J^{2},$ the ratio of the
$C_{J}-$ and $C_{\Delta }-$ originated terms Eq.(5) is of the order
of $C_{J}T_{B}/JC_{\Delta},$ thus containing a small factor of
$T_{B}/J$ that decreases the role of the $C_{J}-$originated term.

 \begin{figure}
\centerline{\epsfig{figure=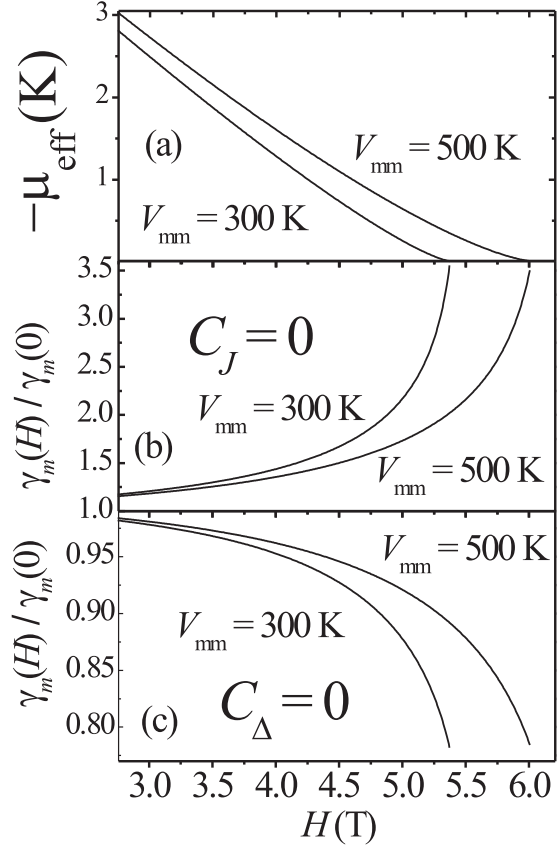,width=10cm}} \caption{Calculated
$\mu_{\rm eff}$ (a) and damping (b,c) for different spin-lattice couplings
as function of field $H$ at $T=3.5$ K. For $H<2.5$ T the behavior of damping
is almost flat, in agreement with the experiment. $\Delta=6.5$ K, $H_g=4.5$
T. The results depend on $V_{\rm mm}$ which determines the $H_c$.}
 \label{fig3}
 \end{figure}

Our approach can qualitatively explain the decrease in the damping rate when
the field exceeds some value corresponding to the broad maximum at $H=9.7$ T
in Fig. 1. The above consideration is based on the low-energy singularity of
the $f_B(\varepsilon)$ increasing as $1/\overline{\varepsilon}$ at $H=H_c$.
At $H>H_c$ the number of quasiparticles $\overline{n}$ rapidly increases
with $H$ \cite{nikuni00}. The spectrum of excitations in this case is
$E(k)=v_{c}k$, with $v_{c}\sim\overline{n}^{1/2}$ being the sound velocity
in the condensate. With the increase of $H$, and, correspondingly
$\overline{n}$, the excitation energy increases, and, therefore, the role of
thermal excitations decreases. Thus, the increase in $H$ drives the
condensate effectively towards the zero-temperature behavior. The system
demonstrates a zero-$T$ behavior when $\overline{n}$ becomes much larger
than $n_{B}$ corresponding to the BEC at given $T$. At zero $T$ the
distribution of the out-of condensate particles which contribute to the
damping at small momenta is $N_{E}(E\rightarrow 0)={mv_{c}^{2}}/{2E(k)}
\cite{Landau58}$. Since $E(k)$ is linear in $k$, the singularity in Eq.(4)
at $\mu_{\rm eff}=0$ becomes weaker, and the damping decreases. To estimate
the field at which the behavior of the condensate is close to the $T=0$
limit we note that at $T>0$ thermal excitations and interaction $V_{\rm mm}$
produce out-of condensate particles. The concentration of the magnons at
which the system demonstrates a crossover to the $T=0$ behavior is
determined by the condition that the concentration of the particles out of
the condensate arising due the interaction ($\sim n\sqrt{na^{3}}$) is larger
than the concentration of thermally excited particles proportional to
$T^{3/2}$, where $a=V_{\rm mm}l^{3}m/4\pi$ is the magnon-magnon scattering
length. In TlCuCl$_3$ $a$ and $l$ are close to each other since the
hard-core boson has a spatial size of the Cu-Cu distance. As a result, the
critical concentration which determines the crossover to zero-temperature
behavior and, in turn, the broad maximum position is $\overline{n}_{\rm
cr}\sim\overline{n}_{B}(l/a\overline{n}_{B}^{1/3})\sim 10\overline{n}_{B}$.
On the other hand from the data on $n(H)-$dependence of Ref. \cite{nikuni00}
one expects that at $T=3.5$ K, $\overline{n}(H=10 {\rm T})$  is an order of
magnitude larger than $\overline{n}_{B}(H_c)$, in a good qualitative
agreement with our estimate.

Following Eqs.(1) and (5) one can estimate $C_{\Delta }$.
The energy density in the sound wave with $\omega =cq$ ($c$ is the sound velocity) is:
$w\sim\kappa u_{0}^{2}{\omega^{2}}/{c^{2}}$,
where $\kappa $ is the elastic constant. The dissipation rate estimated from
Eq.(1) is:
\begin{equation}
\frac{dw}{dt}\sim \frac{F^{2}}{m}\frac{1}{\sigma \left\langle v\right\rangle
}\sim C_{\Delta }^{2}\frac{\omega ^{4}}{c^{4}}u_{0}^{2}\frac{1}{m\sigma
\left\langle v\right\rangle }.
\end{equation}
Therefore, the relaxation rate $(dw/dt)w^{-1}\sim\omega^{2}$.
Let us consider as an example $\gamma_{m}(0)$ that with the
estimates given above can be written as:
\begin{equation}
\gamma _{m}(0)\sim \frac{dw}{dt}\frac{1}{w}\sim \frac{C_{\Delta }^{2}}{\rho
c^{2}}\frac{\omega ^{2}}{c^{2}}\frac{1}{l}
\frac{J}{\sqrt{\Delta T }},
\end{equation}
where we took into account that far from the transition point $\left\langle
1/v\right\rangle \sim 1/\left\langle v\right\rangle $ with $\left\langle
v\right\rangle \sim \sqrt{T/m}$. From the
theoretical $\gamma _{m}(H)-$%
dependence (Fig. 3), we conclude that $\gamma _{m}(H_{g})/$ $\gamma
_{m}(0)\sim 2$. From the experimental data in Fig. 1 we obtain $\gamma
_{m}(H_{g})-$ $\gamma _{m}(0)\sim 0.1$ dB, and, therefore, $\gamma _{m}(0)$
corresponds to a damping of the order of 0.1 dB. For the thickness of the
sample of 2 mm and the sound velocity $c=3\cdot 10^{5}$ cm/s, the pulse
travelling time $t_p$ is 6.6$\cdot 10^{-7}$ s, and, therefore, $\gamma
_{m}(0)\sim 0.2\cdot 10^{4}$ s$^{-1}$. The density $\rho = 5$
g/cm$^{3}$, $\omega \sim 3\cdot 10^{7}$ s$^{-1},$ and $l=\Omega ^{1/3}\approx 5$
\AA, where $\Omega$ is the unit cell volume per chemical formula yield
$C_{\Delta }$ of the order of 200 K.

We presented ultrasonic attenuation experiments as well as a theoretical
analysis on the field-induced changes in the spin dimer system TlCuCl$_{3}$.
These experiments are in agreement with the magnon Bose-Einstein
condensation and summarized as a sharp peak in the attenuation $\alpha(H)$
in the close vicinity of the transition and a second broader maximum at
larger fields, deep in the long-range ordered phase. The peak in $\alpha(H)$
shows a hysteresis crossing the phase boundary with increasing/decreasing
field. This effect is attributed to a first-order contribution to the phase
transition. The exponent $\phi \approx 2$ in the relation $n_{B}\sim T_{B}^{\phi
}$ and the $H$-dependence of sound attenuation originate from the
'relativistic' dispersion of magnons. The spin-lattice coupling is due to
a phonon-induced modulation of the spin gap with the coupling constant
$C_{\Delta }\sim 200$ K. The recent observation \cite{Kodama02,Wolf} of
triplet crystallization in the 2D compound SrCu$_2$(BO$_3$)$_2$ in large
magnetic fields with similar aspects of spin-phonon coupling demonstrate
that these phenomena are of a general nature.

We acknowledge important discussions with D. Lenz, M. Takigawa, B. Wolf and
B. L\"uthi, and financial support by DFG/SPP 1073 and the Austrian Science
Fund project P-15520. Experiments have been performed at the High-Field Laboratory
of the TU Braunschweig with the expert help of R. Hofmann and H.
Simontowski.


\begin{thebibliography}{99}

\bibitem{Auerbach} A. Auerbach, {\it Interacting Electrons and Quantum
Magnetism}, Springer, Graduate Text in Contempoary Physics, New York, 1994.

\bibitem{Lemmens03} P. Lemmens
{\it et al.}, Phys. Reports \textbf{375}, 1 (2003).

\bibitem{oshikawa97}
M. Oshikawa {\it et al.}, Phys. Rev. Lett. \textbf{78}, 1984 (1997).

\bibitem{ultrasound}
R. Joynt {\it et al.},
%and L. Taillefer,
Rev. of Mod. Phys. \textbf{74}, 235 (2002).

\bibitem{pankert90}
J. Pankert  {\it et al.},
%G. Marbach, A. Comberg, P. Lemmens, P. Froening, S. Ewert,
Phys. Rev. Lett. \textbf{65}, 3052-3055 (1990).

\bibitem{willett63}  R.D. Willet {\it et al.},
J. Chem. Phys. \textbf{38}, 2429 (1963).

\bibitem{tanaka96}  H. Tanaka {\it et al.},
J. Phys. Soc. Jpn. \textbf{65}, 1945 (1996).

\bibitem{takatsu97}  K. Takatsu {\it et
al.}, J. Phys. Soc. Japan \textbf{66}, 1661 (1997).

\bibitem{Matsumoto02}  M.
Matsumoto  {\it et al.}, Phys. Rev. Lett. \textbf{89}, 077203 (2002).

\bibitem{Giamarchi99}  T. Giamarchi and A. Tsvelik, Phys. Rev. B
\textbf{59}, 11398 (1999).

\bibitem{nikuni00}  T. Nikuni  {\it et
al.}, Phys. Rev. Lett. \textbf{84}, 5868 (2000).

\bibitem{oosawa99}  A. Oosawa
{\it et al.}, J. Phys.: Condens. Matter \textbf{11}, 265 (1999).

\bibitem{oosawa01}  A. Oosawa {\it et al.},
Phys. Rev. B \textbf{63}, 134416 (2001).

\bibitem{tanaka01}  H. Tanaka {\it et al.},
J. Phys. Soc. Japan \textbf{70}, 939 (2001).

\bibitem{takigawa01} O. Vyaselev {\it et al.},
Physica B, in print (2003)

\bibitem{Cottam74}
M.G. Cottam, J. Phys. C.:\ Solid State \textbf{7}, 2919 (1974).

\bibitem{cavadini01}  N. Cavadini, {\it et al.},
Phys. Rev. B \textbf{63}, 172414 (2001).

\bibitem{oosawa02}  A. Oosawa {\it et al.},
Phys. Rev. B \textbf{65}, 94426 (2002).

\bibitem{takatsu98}  K. Takatsu {\it et
al.}, Journal of Mag. and Mag. Mat. \textbf{177-181}, 697 (1998).

\bibitem{cavadini02}  N. Cavadini {\it et al.},
Phys. Rev. B \textbf{65},  132415 (2002).

\bibitem{Landau58}  L. D. Landau and E. M. Lifshitz, {\it Statistical Physics},
London, Pergamon Press, 1958, 484 p.

\bibitem{Kodama02}  K. Kodama {\it et
al.}, Science \textbf{298}, 395, (2002).

\bibitem{Wolf}
B. Wolf  {\it et al.}, Phys. Rev. Lett. {\bf 86}, 4847 (2001).

\end{thebibliography}
\end{document}